# Direct Observation of Energy Transport Dynamics and High Thermal Conductance across Single Solid–Molecule Junctions

Md. Shahriar H. Shuvo, Xing He, Mithun Ghosh, Ding-Shyue Yang*

*Department of Chemistry, University of Houston, Houston, Texas 77204 United States*

*To whom correspondence should be addressed. Email: yang@uh.edu

**Abstract**

Interfaces play a crucial role in energy transport at the nanoscale. However, direct experimental observations of interfacial thermal conductance across molecular junctions have remained challenging due to the high spatiotemporal resolution required for probing. Here, we report dynamic energy transport processes across multi-component molecular junctions observed at the atomic level by employing reflection ultrafast electron diffraction. A clear temporal sequence of energy transfer is revealed at early times following photoexcitation of Au(111) surfaces chemically bonded with self-assembled monolayers (SAMs) of alkanethiols: from the gold surface layer (SL) to the head groups of a SAM and then to the $CH_2$–$CH_2$ methylene lattice. Remarkably, the structural dynamics of the gold SL differ significantly from those of clean gold. Furthermore, the methylene lattice dynamics exhibit chain-length insensitivity but with a length-dependent retention time to reach full thermalization. Quantitatively, we find an agreement in the increased out-of-plane atomic motions between the surface gold atoms and the SAM methylene units, signifying the nature of motion-based coupling for interfacial energy transport. High interfacial thermal conductance (~300 MW $m^{-2}$ $K^{-1}$) and high thermal conductivity for the methylene lattice (~60 W $m^{-1}$ $K^{-1}$) are obtained under the condition of impulsive heating, which provide strong support for previous room-temperature theoretical predictions and also the ballistic nature of intrachain heat transfer. This time- and spatially-resolved experimental approach will enable future quantitative assessment of interfacial energy transport across solid–molecule junctions.

Energy transport and coupling at the nanoscale has been under intense research for nanosized structures and contacts and for molecular junctions (*1–6*). As modern devices incorporate dissimilar materials and their dimensions are comparable or below the electron/phonon mean free paths, the interfaces often play a crucial role in the electronic (*7–9*) and vibrational/thermal transport phenomena (*1, 10*). For the latter, experimental studies, especially those using transient thermoreflectance methods, has provided evidence for enhancements of interfacial thermal conductance by chemical bonding (*11–13*), a bridging effect to mediate the vibrational spectrum mismatch (*14*), or cooperative behavior between packed molecules (*15*). These ensemble-based results illustrate the use of a self-assembled monolayer (SAM) of organic molecules as an intermediate layer, whose bonded or van der Waals (vdW) interactions with metal or dielectric surfaces provide an adaptable platform to adjust thermal and electrical transport across such molecular junctions (*16, 17*). On the theoretical side, atomic-level molecular dynamics and *ab initio* simulations beyond the continuum-based mismatch models have gained significance to examine thermal boundary conductance (TBC) in detail for effects such as the transport mechanism, chain-length dependence, impact of interfacial coupling strength, etc. (*1, 18, 19*). To date, the influence of substrate–molecule interactions and ordering of molecular assemblies on interfacial thermal conductance, with the opportunity for active control by external stimuli, is still an open question in the area of molecular phonon engineering (*5*).

Major experimental challenges lie in the lack of direct observations of the dynamic energy transport processes across multi-component systems at the atomic/molecular level. First, almost all existing experimental methods report phenomenological results without atomic details. For example, the derivation of interfacial thermal conductance from often-used time-domain thermoreflectance (TDTR) measurements relies on the assumption of accurately known thicknesses, specific heats, and thermal conductivities of the metal transducer and involving solid(s)/liquid, as well as the view of the SAM layer as one abrupt interface in the thermal model (*1, 11, 20, 21*). However, the derived TBC values do not reveal the underlying processes directly. Second, most measurements were actually made across *two* molecular junctions—that is, both contacts of a mediating SAM layer or molecule(s)—instead of giving results for a specific side. An issue has been raised regarding the uncertainties in the contact conditions and defects given the challenge in the fabrication of sandwiched structures (*5, 15, 22*). Third, a substrate-supported SAM consists of multiple components with potentially different phonon behaviors. In fact, even for clean Au(111), a direct structure-probing study has shown additional transient motions formerly unnoticed for surface gold atoms compared to the bulk following photoexcitation (*23*). Therefore, it is reasonable to also ask how chemical bonds and lateral vdW interactions affect impulse-induced atomic motions in the Au–SAM system. To date, various experimental (*4*, *24*, *25*) and theoretical (*18*, *26*, *27*) studies have indicated a length insensitivity of room-temperature thermal transport through the alkane chains after a sufficient number of methylene units, although the experimental TBC values have been mostly much lower than predicted by theories (*26*, *28*). A closer agreement is seen for a length-independent TBC value of 220±100 MW m$^{-2}$ K$^{-1}$ obtained from the time-resolved sum-frequency generation (trSFG) measurements, made with impulsive back-illumination of supported gold films bonded with *n*-alkanethiol SAMs on the front side and monitoring the terminal methyl groups (*29*). However, at the impulsive temperature jump of ~35 K used (*30*), the physical picture given by the molecular simulations in Ref. (*29*) is not adequate.

Here, we report direct observations of impulse-initiated energy transport dynamics of single-side Au(111)–SAM molecular junctions using ultrafast electron diffraction (UED) in reflection geometry. The direct structure-probing with subatomic-level details is further enabled by tracking the time-dependent changes of different diffraction features originated from the different moieties of the composite system (*31*). We find that the chemical bonding and vdW packing of thiolate molecules to the surface gold atoms drastically alters the photoinduced structural dynamics of the Au(111) surface layer (SL), on both ultrashort and long timescales. A temporal order in the coupling of atomic motions is established for the early-time energy transport from the gold SL to the head-group (HG) region and then to the methylene chains, and in the reversed direction with thermalization and recovery at longer times. It is further found that the Au–HG interface acts as the main contributor to the mild interfacial thermal resistance, whereas length insensitivity for the fast thermal conductance across the methylene lattice is confirmed but with a length-dependent retention time prior to recovery. By using a thermal model, we obtain an effective TBC value in agreement with the theoretical predictions in the literature and a through-chain thermal conductivity resembling that of crystalline polyethylene (PE). Moreover, a transient ordering of the interfacial system is found during thermal relaxation on longer times, which signifies the impact of collective motions of the vdW-packed chains on chemically-bonded HGs and surface gold atoms.

**Dynamics of multi-component molecular junctions probed by reflection UED**

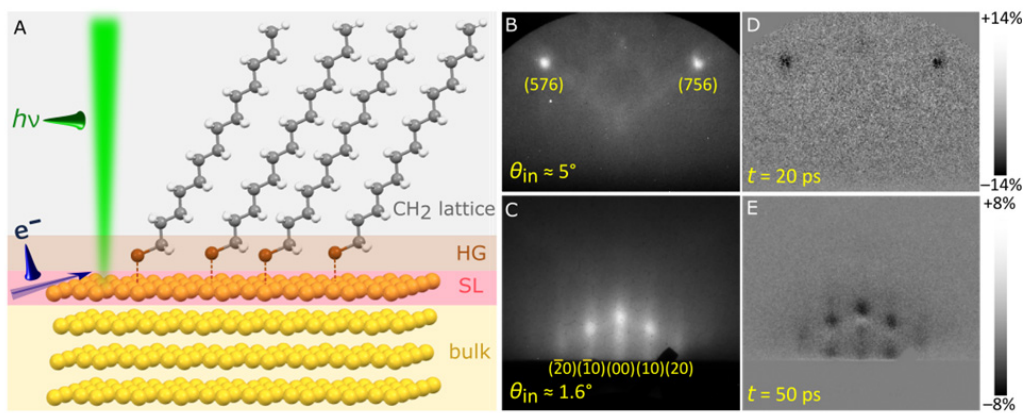

**Fig. 1. Schematic and diffractions of the Au(111)–SAM system probed by reflection UED.** (A) A monolayer of *n*-alkanethiolates with 14-carbon chains chemically bonded to Au(111) (carbon in gray, hydrogen in white, sulfur in brown, and gold in yellow). An impulsive laser excitation of the gold substrate is made by 515 nm and the pulsed electron beam propagates parallel to $[10\bar{1}]$ of Au(111). (B and C) Electron diffraction images acquired from an as-deposited specimen at the select incidence angles of ~5.0° and ~1.6°, respectively. (D and E) Diffraction differences obtained at 20 ps and 50 ps referenced with the corresponding negative-time frames in B and C. Diffraction intensity decrease is seen as darkness with the scale bar given on the right.

Shown in Fig. 1A and fig. S1 is the molecular model of a SAM of *n*-alkanethiolates bonded to Au(111), which is composed of the alkane chains packed by the vdW interactions, the sulfur HG region, the gold SL, and the gold bulk underneath. The latter two need to be separately considered according to the recent discovery that the atoms of clean gold surfaces exhibit different structural dynamical behavior compared to the bulk within first nanosecond (ns) following laser excitation (*23*). Different components of the composite system can be specifically probed by changing the incidence angle $\theta_{in}$ of the grazing electron beam and observing the corresponding diffraction features (*31*). Specifically, at lower $\theta_{in}$ = 1.4°–2.0°, Bragg spots are observed as a result of the three-dimensional (3D) $CH_2$–$CH_2$ methylene lattice formed by the assembled all-trans chains, together with the overlaying weak diffraction streaks given by the HGs that essentially form a $(\sqrt{3} \times \sqrt{3})R30°$ two-dimensional (2D) sublattice (Fig. 1C). The entire SAM thickness is probed based on the estimated probe depth of ≤3 nm by $l \cdot \sin\theta_{in}$ where $l \sim 97$ nm is the elastic mean free path of 30-keV electrons. However, at a high incident angle of $\theta_{in} \sim$ 5.0°, electrons penetrate the molecular layer and access the surface gold atoms to yield well-defined Bragg spots (Fig. 1B) (23). Thus, these distinct diffraction features enable us to directly probe the structural and energy transport dynamics across the multi-component solid–molecule junction with high spatiotemporal resolution. Following the impingement of 515-nm light at an apparent fluence of 4.4 mJ/cm², the initial photoexcitation of Au(111) leads to diffraction changes observed in the subpicosecond (sub-ps) to ns temporal range (Fig. 1, D and E, referenced with the negative-time frames). The notable intensity decreases of different diffraction features signify increases in the respective atomic motions. To quantify the changes, the horizontal intensity profiles across the centers of the spots and streaks are fitted with a Lorentzian function.

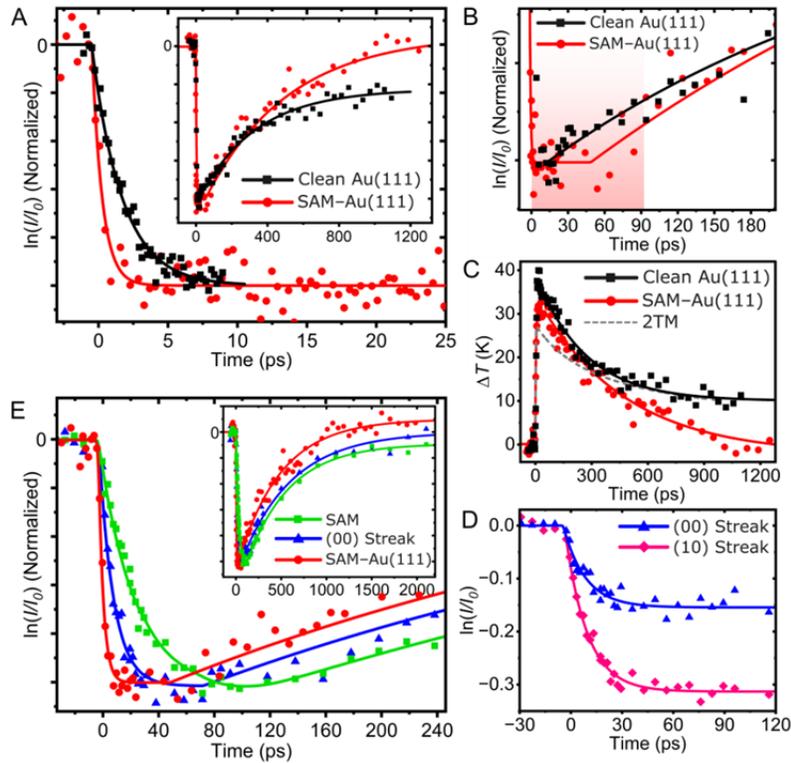

**Fig. 2. Time-resolved structural dynamics of the Au(111)–SAM system at 4.4 mJ/cm².** All solid lines are guides to the eye. (A) Comparison of the early-time diffraction intensity changes acquired from clean (black) and SAM-bonded (red) Au(111) SLs. The inset shows the recoveries on a longer timescale. (B) A magnified view of the inset of A around the maximum decreases. The colored region indicates the retention and delayed recovery of the diffraction change for the SAM-bonded Au(111) SL compared to that of clean gold. (C) Comparison of the photoinduced SL temperature increases and the bulk temperature theoretically given by 2TM (gray dashed line). (D) Diffraction intensity changes of the center (00) (blue) and side (10) (pink) streaks. (E) Comparison of the time-dependent diffraction intensity changes originated from the gold SL (red), the HGs (blue), and the methylene lattice (green). The inset shows the longer timescale.

## Ultrafast thermalization of chemically-bonded surface gold atoms

We first examine the photoinduced responses of the chemically-bonded Au(111) SL (Fig. 1, B and D), which absorbs photons and becomes the initial source of energy and increased atomic motions across the junction (Fig. 2, A to C). Surprisingly, three behavioral differences are evident compared to the structural dynamics of clean Au(111) (*23*). First, at early times, instead of a Bragg intensity decrease with an apparent time constant of 2.2 ps (Fig. 2A, black), which is sufficiently explained by the two-temperature model (2TM) (*32*), the SAM-bonded Au(111) SL exhibits a sub-ps dynamical response close to the instrumental response time (Fig. 2A, red). This signifies an accelerated electron–phonon (*e*–ph) thermalization at the SL as a result of the different chemical environment and forces, possibly via electronically-coupled processes (*33*, *34*). We note that the observable here is directly lattice structure-based and related to the outcome of *e*–ph coupling, not electronically-dominant as given by most all-optical methods. Second, the reduced diffraction intensity remains at a constant level for about 50 ps before starting to recover, which is distinctly different from the behavior of an immediate recovery for clean Au(111) SL (Fig. 2B). Third, the diffraction intensity of the SAM-bonded Au(111) SL is nearly fully recovered within ~1 ns, whereas clean Au(111) shows a remainder of the diffraction reduction by ~27% in the same time window (Fig. 2A, inset).

The time-dependent effective temperatures $\Delta T$ of the Au(111) SL are derived based on the maximum Bragg intensity decrease of 17 % and 14 % at 4.4 mJ/cm$^2$ for clean and SAM-bonded gold, respectively (Fig. 2C and fig. S2); the surface atoms exhibit additional transient out-of-plane mean-square displacements (MSDs) compared to the bulk described by 2TM (Fig. 2C, dashed line) (*23*). We find that the SAM-bonded Au(111) SL yields an initial jump of $\Delta T$ ~ 30 K comparable to that of ~37 K for clean gold. The level of energy impulse used in this study is therefore moderate and does not cause laser-induced damages to the densely packed SAMs.

## Structural and energy transport dynamics of SAMs

Now we move our attention to the time-dependent intensity changes of the diffraction streaks and SAM diffraction spots (Fig. 1, C and E), which give the structural dynamics of the bonded HGs and the methylene lattice of *n*-alkanethiolates, respectively (Fig. 2E). The temporal order of the diffraction changes is clear, where those associated with the sulfur HGs trail those from the Au(111) SL but precede those of the methylene lattice, for both the rise and recovery parts. Such a sequence of dynamical changes demonstrates the unique capability of UED in reflection geometry to study interfacial junctions. It is understandable that increases in the HG and $CH_2$ atomic motions (which cause diffraction intensity decreases) are anticipated as a result of upward energy transport from the supporting surface gold atoms. At later times following the thermalization, thermal dissipation in the reverse direction takes over as the laser-heated Au(111) bulk and SL continue to recover. Therefore, quantitative examinations of these changes over the observed temporal window are helpful to elucidate the underlying dynamical processes.

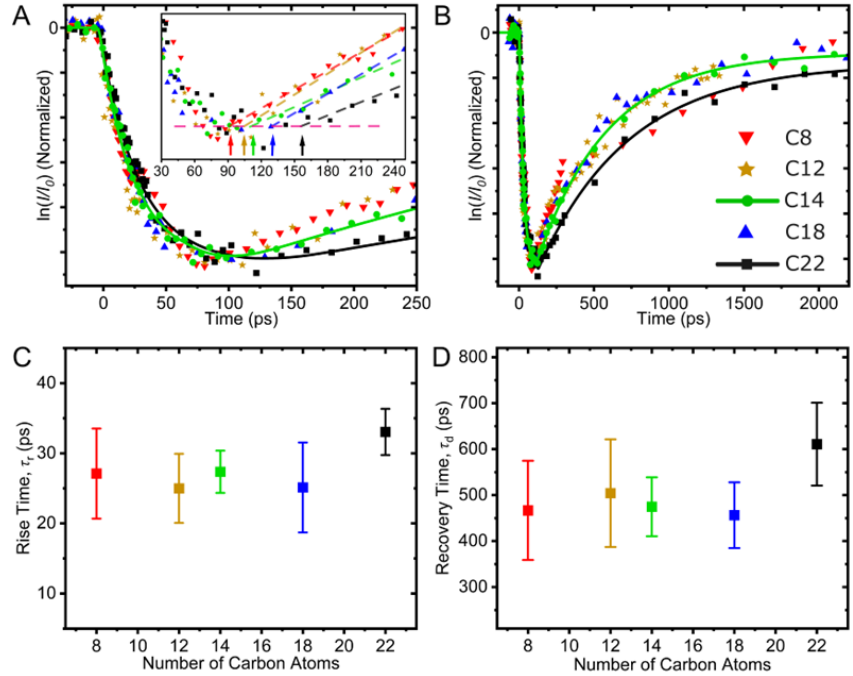

**Fig. 3. Time-resolved changes of the Bragg diffraction intensities from the methylene lattice for different chain lengths.** Solid curves are guides to the eye. (A) Similar early-time rises for the intensity decrease. The inset shows a magnified view near the maximum decrease, with the arrows indicating the length-dependent retention times deduced from the intersections of linear fits. (B) Similar time-dependent recoveries observed at longer times. The legend gives the color-coded symbols used in all panels for the corresponding chain lengths (i.e., carbon numbers). (C and D) The apparent rise and recovery time constants for different chain lengths, extracted from the fits with simple exponential functions. The error ranges indicate the 95% confidence level.

The increased atomic motions are derived according to the Debye–Wallar model (35),

$$\ln\left(\frac{I(t)}{I_0}\right) = -4\pi^2 s^2 \cdot \Delta\langle u^2(t)\rangle_{\vec{s}} \quad (1)$$

where $I_0$ and $I(t)$ are, respectively, the diffraction intensities of a diffraction feature before the zero of time and at time $t$ after excitation, $s$ is the momentum transfer, and $\Delta\langle u^2(t)\rangle_{\vec{s}}$ is the averaged increment of MSDs projected along the momentum transfer direction $\vec{s}$. From the intensity decreases of the center and side streaks being ~15% and ~31%, respectively (Fig. 2D), we estimate the in-plane horizontal MSDs of the HGs to be about 5.4 times of their out-of-plane vertical MSDs (see supplementary text 1 for details). Intriguingly, such a large ratio of the photoinitiated dynamical motions along different directions resembles the ratio of 5.2 for the MSDs along the vdW interchain axes vs. the intrachain methylene direction measured from crystalline PE in thermal equilibrium (*36*). This comparison signifies the importance to take into account the prominent anisotropy in both the chemical interactions (vdW vs. bonding) and the structural packing (between vs. along the methylene chains) of a molecular assembly when examining its dynamical behavior.

Shown in Fig. 3 are the structural dynamics probed from the methylene lattice for different chain lengths of $n$-$C_nH_{2n+1}SH$ (Cn) SAMs with n = 8, 12, 14, 18, and 22. It is evident that all of the SAMs studied exhibit a similar apparent rise time (Fig. 3, A and C) and comparable recovery (Fig. 3, B and D), with a certain length-dependent maturation time (Fig. 3A, inset). The former length insensitivity of the overall dynamical behavior is consistent with prior experimental and theoretical reports. However, the latter length dependence is likely the result of energy transport and retention in the entire probed methylene lattice that contributes to the diffraction signals (see supplementary text 2 and fig. S3). Importantly, the time-resolved changes of the center and side streaks show essentially the same rise time of ~13 ps and retention time, both independent on the chain length (fig. S4). Such an observation is consistent with the physical picture that the chemically-bonded HGs experience increased motions and receive energy from the photoexcited surface gold atoms, and hence they exhibit largely the same initial dynamics uninfluenced by the methylene lattice above them if the chain length is sufficient. It is also worth noting that the length-dependent retention times extrapolate to an intercept of about 52±7 ps for no methylene spacer group (fig. S5), which is consistent with the aforementioned retention period as a plateaue of the diffraction change for the Au(111) SL (Fig. 2B, red).

It is informative to compare the impulse-induced atomic motions of different moieties of the Au(111)–SAM system, which are along the surface normal direction given the reflection probing geometry. We obtain an MSD increase of $3.20 \times 10^{-3}$ $Å^2$ for the methylene lattice from the Bragg intensity decrease by ~8%, which signifies an average increase in the out-of-plane root-mean-square displacement (RMSD) by ~0.0084 Å (see supplementary text 3 and fig. S6). For the gold SL, the vertical RMSD is estimated to be ~0.0088 Å based on the temperature jump of ~30 K (Fig. 2C) and the Debye model (*35*),

$$\langle u_\perp^2(T) \rangle = \frac{3\hbar^2}{mk_B\Theta_D}\left[\frac{1}{4} + \left(\frac{T}{\Theta_D}\right)^2 \int_0^{\Theta_D/T} \frac{x\,dx}{e^x - 1}\right] \cong \frac{3\hbar^2 T}{mk_B\Theta_D^2} \quad (2)$$

where $\hbar$ is the reduced Planck constant, $m$ is the atomic mass of gold, $k_B$ is the Boltzmann constant, and $\Theta_D$ = 83 K is the Debye temperature of the gold surface (*37*). The close agreement between these two RMSDs signifies a motion-based coupling for interfacial energy transport, which has also been seen in surface-supported, vdW-separated molecular thin-film systems under impulsive heating (*38*, *39*). Here, however, strong chemical bonds connect between the supporting surface, the HGs, and the methylene lattice, and a 3D order is present in the SAM structure. Thus, the dynamical responses observed in Fig. 2E are faster compared to those of the molecular thin films without a good interlayer stacking order (*38*, *39*).

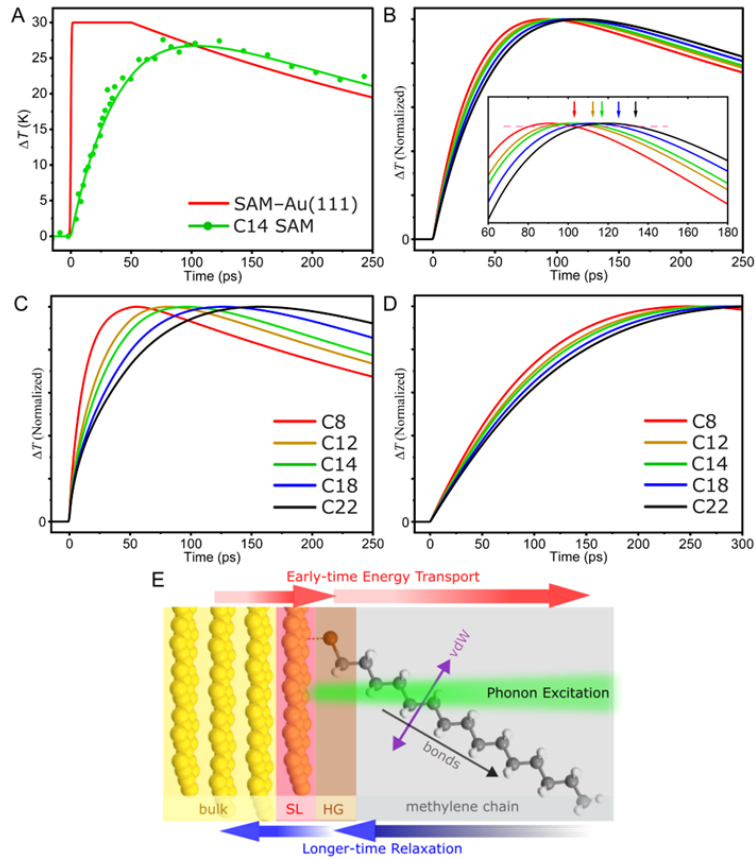

**Fig. 4. Time-dependent dynamics of the methylene lattice assessed by a thermal model.** (A) Comparison of the experimentally-deduced temperatures of a C14 SAM (green dots) and the theoretical time-dependent curve obtained with the best $\sigma$ and $K$ values (green line). The red curve shows the experimentally-determined temperature profile used for the gold SL. (B to D) Theoretical curves of temperature evolution with the length dependence obtained with (B) $\sigma$ = 300 MW m$^{-2}$ K$^{-1}$ and $K$ = 60 W m$^{-1}$ K$^{-1}$, (C) $\sigma$ = 300 MW m$^{-2}$ K$^{-1}$ and $K$ = 0.1 W m$^{-1}$ K$^{-1}$, and

(D) $\sigma = 60$ MW m$^{-2}$ K$^{-1}$ and $K = 60$ W m$^{-1}$ K$^{-1}$. The inset in B shows a magnified view around the maximum temperature changes, with the arrows indicating the onset times for clear recovery taken as 0.5% below the maximum. (E) Schematic of impulse-induced thermal transport and relaxation in a single-sided Au–SAM molecular junction at early (red arrows) and longer (blue arrows) times, respectively.

**Thermal boundary conductance model**

We use a thermal model to examine the observed time-resolved behavior. Each CH$_2$ unit is associated with an effective temperature to describe its increased MSD; a function with an instantaneous 30-K jump, a 50-ps retention duration, and an exponential recovery is used for the gold SL temperature (Fig. 4A, red). The coupling between the gold SL and the first CH$_2$ of the alkane chain via the HGs is treated by TBC ($\sigma$), whereas the thermal transport along the chain is modeled by the one-dimensional (1D) diffusion equation with phononic thermal conductivity ($K$) (see supplementary text 4). Given that the SAM thicknesses are within the probe depth of reflection UED, a weighted average of the temperatures of all methylene units is used to compare with the experimentally-deduced results (fig. S3). We find that both $\sigma$ and $K$ need to be large to reproduce the UED results of length insensitivity for the early-time rise of the SAM temperature; the length-dependent onset time for noticeable recovery is generally reproduced by the model also (Fig. 4B). In a test scenario, if a lower $K$ of the order of 0.1 W m$^{-1}$ K$^{-1}$ is used considering that of disordered PE (*40*, *41*), a clear length dependence is resulted even when the TBC value is kept high (Fig. 4C). Such a result is consistent with high thermal conductivity of the order of $10^1$–$10^2$ W m$^{-1}$ K$^{-1}$ for highly-ordered PE along the well-packed methylene chains (*42*, *43*) as well as the ballistic nature of this molecular thermal transport (*5*, *29*, *44*). In another scenario, if the experimentally-determined value of ~60 MW m$^{-2}$ K$^{-1}$ (*11*) is used for $\sigma$ with the high $K$, the SAM temperature rise becomes too slow although the length-insensitivity feature is kept (Fig. 4D). Thus, our UED study yields $\sigma \sim 300$ MW m$^{-2}$ K$^{-1}$ and $K \sim 60$ W m$^{-1}$ K$^{-1}$ to obtain a theoretical curve in good quantitative agreement with the UED observation (Fig. 4A, green), which signifies high TBC and high intrachain thermal conductance of this molecular junction, at least under the condition of impulsive heating. A temporal sequence of the energy transfer across different moieties at early times is depicted in the upper part of Fig. 4E.

**Transient structural ordering of the Au–SAM junctions**

Following thermalization in the SAM and the relaxation of the gold SL already starting at ~50

ps, the excess heat in the molecular junction dissipates in the reversed direction (Fig. 4E, blue arrows). At 2 ns, a residual temperature increase of 7 K remains in bulk gold based on 2TM (Fig. 2C, black). On this long timescale, it is found that the recovery dynamics of the SAMs do not significantly depend on $\sigma$ and $K$, although a better match is found with them at high values (fig. S7). Therefore, we argue that the confirmation of the large thermal conductance values, which may still be valid for slower processes, requires experimental methods with high surface sensitivity and spatial resolution such as that shown in the present work.

Interestingly, we find that the gold SL exhibits a full recovery in 1.2 ns, significantly faster than the bulk even with the initial ~50-ps retention time (Fig. 2A, inset, and Fig. 2C). In fact, the gold SL diffraction intensity even sees an enhancement of the peak intensity by ~10% at ~2 ns compared to that prior to photoexcitation, whereas the (00) streak diffraction from the HGs also returns to the original level (Fig. 2E, inset, red and blue). Furthermore, a slight decrease in the gold SL diffraction width is also observed at long times (fig. S8). With the understanding that a small amount of residual heat still remains in the Au–SAM interfacial system during this time, these observations signify the presence of transient structural ordering that boosts the diffractions, which was previously suggested in an early reflection UED study of a disordered assembly of very short thiolates on gold (*45*). A small-angle transmission UED study also describes the likelihood of dynamic annealing for the observed enhanced homogeneity of the supracrystals of dodecanethiol-capped gold nanoparticles (*46*). We note that a certain extent of disorder is unavoidable in the deposited Au–SAM system (*16*, *47*, *48*). Hence, cooperative small nudges of the vdW-packed chains dynamically alleviate the intrinsic structural defects (and, as a result, boost a small increased order) in the supposedly all-trans conformation of the methylene lattice, the sulfur-HG adsorption sites, and the bonded gold SL atoms, whose effect is equivalent to a reduced temperature by several degrees. However, this transient ordering disappears when the system is fully relaxed.

**Summary and outlook**

This time-resolved study of single Au–SAM junctions by reflection UED reveals the previously unnoticed complex structural and energy transport dynamics at the atomic level following impulsive photoexcitation and heating of the gold surface. The significantly different behavior of the surface gold atoms on ultrashort and long timescales results from the coupling to the chemical bonded HGs and the collective influence of the vdW-packed alkane chains. In such a system, the thermal boundary resistance is experimentally confirmed to be low with chain-length insensitivity, although the methylene lattice with high thermal conductivity takes additional time to thermalize for longer chains. Our findings indicate the mutual impacts of the supporting surface and the adsorbate assembly on their own dynamical responses, which point out the need to not only examine the adsorbate but also scrutinize the structural dynamics of a supporting surface in a solid–molecule system. With the atomic-level spatiotemporal resolution, it is now crucial to study different molecular junctions to investigate the effects such as packing orders, forces, and structural inhomogeneity of molecular assemblies, surface chemical and electronic interactions, and functionalization, etc. on interfacial dynamics.

## ACKNOWLEDGEMENTS

We thank N. Chiang for helpful discussions. **Funding:** The support by the National Science Foundation (CHE-2154363) and the R. A. Welch Foundation (E-2263) is acknowledged. **Author contributions:** D.-S.Y. conceived the idea and supervised the work; Md.S.H.S., X.H., and M.G. conducted the measurements; Md.S.H.S. and X.H. performed the analysis; D.-S.Y. and Md.S.H.S. interpreted the results and wrote the manuscript. All authors contributed to the review and editing of the manuscript. **Competing interests:** The authors declare no competing financial interests. **Data and materials availability:** All data needed to evaluate the conclusions in the paper are present in the main text or the supplementary materials.

## SUPPLEMENTARY MATERIALS

Materials and Methods; Supplementary Text; Figs. S1 to S9; References (49–55)

# Supplementary Materials for

**Direct Observation of Energy Transport Dynamics and High Thermal Conductance across Single Solid–Molecule Junctions**

Md. Shahriar H. Shuvo *et al.*

Corresponding author: Ding-Shyue Yang, yang@uh.edu

## Materials and Methods

Single crystals of Au(111) with a surface orientation accuracy of <0.1° (Princeton Scientific) were used to obtain large areas of highly-ordered self-assembled monolayers (SAMs). Prior to the SAM formation on Au(111), the crystals were cleaned by performing several cycles of 20-minute $Ar^+$ sputtering at 1.0–1.5 keV at room temperature followed by 20-minute annealing at approximately 750 K under vacuum with a base pressure below $4\times10^{-8}$ torr. The atomically flat Au(111) surfaces were confirmed by the characteristic diffraction streaks without transmission-like spots observed by reflection high-energy electron diffraction at low incidence angles. The cleaned gold substrates were then removed from the vacuum chamber and immediately used in the solution-based SAM fabrication by immersing the gold substrates in 1-mM alkanethiol solutions in ethanol for 24 hours. The chemicals 1-octanethiol, 1-dodecanethiol, 1-tetradecanethiol, 1-octadecanethiol, and 1-docosanethiol (98%) were obtained from Sigma Aldrich and used directly without further purification. After the SAM formation, the samples were thoroughly rinsed with ethanol, blow-dried with pure nitrogen, and subsequently loaded into vacuum. The total ambient exposure time during rinsing, drying, transporting, and loading was kept below 30 minutes. Measurements of the prepared samples were made in ultrahigh vacuum (UHV) with a base pressure of $\sim2\times10^{-10}$ torr.

Details of the reflection ultrafast electron diffraction (UED) apparatus have been reported (*23*, *49*, *50*). Briefly, a Yb:KGW regeneratively amplified laser system was operated at a repetition rate of 10 kHz and delivered a fundamental output of 1030 nm with a pulse duration of 170 fs. Second harmonic generation (SHG) was employed to produce 515-nm (2.41 eV) pulses from the fundamental beam, which passed through a beam-front tilt setup and were used to photoexcite the gold substrate, not the SAMs. A fraction of the 515-nm beam was further frequency-doubled via another stage of SHG to produce ultraviolet (257 nm) pulses for generation of the photoelectron pulses accelerated to 30 keV from a $LaB_6$ emitter tip. Recorded by an intensified CMOS camera, electron diffraction images of the gold surface layer (SL) were obtained at a higher incidence angle of $\theta_{in} \sim 5°$, whereas those acquired at lower angles ($\theta_{in} \sim 1.8°$) contained diffraction spots corresponding to the assembled three-dimensional methylene lattice accompanied by streak patterns arising from the two-dimensional sublattice of the head groups (HGs). The temporal resolution of the reflection UED apparatus can reach a sub-ps level with the use of a reduced number of <100 electrons per pulse, which was applied for probing early-time ultrafast dynamics of the gold SL. However, a higher number of several hundred of electrons per pulse were typically used to achieve an improved signal-to-noise ratio within a reasonable acquisition time, resulting in an instrumental response time of ~2 ps for the data acquired on the ps to ns timescales. The laser fluence was estimated based on the full widths at half maximum of the laser footprint (180×1640 µm²) on the specimens, which was sufficiently larger than the focused electron footprint (15×170-to-480 µm²) to ensure a uniformly photoexcited region.

## Supplementary Text

**1. Anisotropic structural dynamics of the SAM HGs and crystalline polyethylene**

We use the maximum intensity decreases of the center (00) and side (10) streaks, 15% and 31% respectively (Fig. 2D), to calculate the in-plane and out-of-plane components of the impulsively induced motions of the HGs. The latter can be found from the (00) streak to be $\Delta\langle u^2(t)\rangle_{\text{out}} = 0.013$ Å² for the increased out-of-plane mean-square displacement (MSD) based on the Debye–Waller model (Eq. 1) with the vertical momentum transfer $s_{\text{out}} = 0.550$ Å$^{-1}$. The momentum transfer of the (10) streak contains both the out-of-plane (vertical) and in-plane (horizontal) components. Therefore, the intensity decrease $I/I_0$ can be expressed as

$$\ln(I/I_0) = -0.31 = -4\pi^2(s_{\text{out}}^2 \cdot \Delta\langle u^2(t)\rangle_{\text{out}} + s_{\text{in}}^2 \cdot \Delta\langle u^2(t)\rangle_{\text{in}}),$$

and we obtain the in-plane MSD of $\Delta\langle u^2(t)\rangle_{\text{in}} = 0.070$ Å². The ratio of the in-plane to out-of-plane MSDs is 5.4, which signifies highly anisotropic motions of the HGs in a dynamical setting following ultrafast impulsive heating. This appears to resemble the previously reported MSDs of crystalline polyethylene (PE) derived from x-ray diffraction (XRD) data measured in equilibrium at room temperature, where the MSD along the methylene chain is $U_{cc} \sim 0.017$ Å² and those along the van der Waals (vdW) interchain axes are $U_{aa} \sim U_{bb} \sim 0.089$ Å² based on Fig. 3 of Ref. (*36*). The corresponding lateral-to-intrachain MSD ratio is 5.2. Although measured in different experimental conditions (UED dynamically vs. XRD in equilibrium; the HGs of a SAM vs. crystalline PE), the resemblance of the MSD ratios suggests the level of anisotropic motions expected in chemically-bonded chains packed by vdW interactions. Given that the force constant of a chemical bond is appreciably higher than those effective values for interchain vdW forces, anisotropic MSDs measured at the same temperature and their dynamical increases are expected. This behavioral similarity supports the use of the PE MSD results to determine the MSD of the

SAM methylene lattice at room temperature and the corresponding effective Debye temperature (see below).

2. **Weighted calculation of the effective assembly temperature**

The relatively low scattering cross sections of the carbon and hydrogen atoms result in the entire SAM's contribution to the observed diffraction signals. However, from the viewpoint of elastic scattering, the top carbon layer of a SAM contributes to the outgoing diffraction beam slightly more than the lower carbon layers because of the gradual attention of the penetrating beam (Fig. S4A). Hence, it is important to consider the depth dependence of the diffraction contribution of a SAM. The elastic scattering cross sections of carbon ($\sigma_C$) and hydrogen ($\sigma_H$) are $2.2357 \times 10^{-22}$ m² and $0.1014 \times 10^{-22}$ m², respectively, for 30-keV electrons (*51*). Based on the packing density of the chains ($N$) and therefore the volume ($V$) for two CH$_2$ units, defined by the chain area (21.6 Å²) and unit-cell height (2.2 Å), the electron mean free path is given by

$$l = \frac{1}{N\sigma} = \left(\frac{2\sigma_C + 4\sigma_H}{V}\right)^{-1} \cong 97 \text{ nm}.$$

Therefore, the diffraction probe depth is ~3.0 nm by $l \cdot \sin\theta_{in}$ with the incidence angle of $\theta_{in} = 1.8°$. Considering an exponential attenuation for the original intensity of the electron beam $i_0$ as a function of the depth $z$, i.e. $i(z) = i_0 \exp(-z/l \sin\theta_{in})$, the penetration factor of electrons passing each methylene layer (with $\Delta z = 1.1$ Å being the height) is approximately $\gamma = \exp(-\Delta z/l \sin\theta_{in}) \cong 1 - \Delta z/l \sin\theta_{in} = 0.964$. The diffraction contribution of the $j$-th methylene layer from the top is then $\alpha \gamma^{j-1} i_0$ with $\alpha$ being the diffraction efficiency factor. Thus, the corresponding normalized weight is

$$P_j = \frac{\alpha \gamma^{j-1} i_0}{\sum_m \alpha \gamma^{m-1} i_0} = \frac{(1-\gamma)\gamma^{j-1}}{1-\gamma^n}$$

where n is the total methylene number (Fig. S4B). It is therefore reasonable to use a weighted average when considering an effective temperature for the whole methylene chain to compare with the UED results.

3. **Estimation for the atomic motions and effective Debye temperature of the methylene lattice along the probed surface normal direction**

We use the XRD data of crystalline PE at room temperature ($T_0 = 290$ K) (*36*) to estimate the

MSDs of a SAM along the intrachain and interchain vdW directions, which are $\langle u_\parallel^2(T_0)\rangle = 0.017$ Å$^2$ and $\langle u_\perp^2(T_0)\rangle = 0.089$ Å$^2$, respectively. Because the grazing electron beam probes atomic displacements along the surface normal (out-of-plane) direction (Fig. S6), the effective MSD along the out-of-plane direction prior to impulsive heating is

$$\langle u^2(T_0)\rangle_{out} = \left(\sqrt{\langle u^2(T_0)\rangle_{out}}\right)^2 = \left(\sqrt{\langle u_\parallel^2(T)\rangle}\cdot \cos 30°\right)^2 + \left(\sqrt{\langle u_\perp^2(T)\rangle}\cdot \sin 30°\right)^2$$

which gives a root-mean-square displacement (RMSD) of $\sqrt{\langle u^2(T_0)\rangle_{out}} = 0.19$ Å$^2$. The observed out-of-plane MSD increase $\Delta\langle u^2\rangle_{out} = 3.20\times 10^{-3}$ Å$^2$ means the total MSD at the elevated temperature as $\langle u^2(T)\rangle_{out} = \langle u^2(T_0)\rangle_{out} + \Delta\langle u^2(\Delta T)\rangle_{out}$ and therefore an RMSD increment as

$$\Delta\langle u\rangle_{out} = \sqrt{\langle u^2(T)\rangle_{out}} - \sqrt{\langle u^2(T_0)\rangle_{out}} = 0.0084 \text{ Å}$$

For the gold SL with a surface Debye temperature of 83 K (*37*), the out-of-plane MSD at room temperature is $\langle u^2(T_0)\rangle = 0.0311$ Å$^2$ based on Eq. 1. It becomes $\langle u^2(T)\rangle_{out} = 0.0343$ Å$^2$ after an impulsive temperature jump of ~30 K (Fig. 2C, red). The corresponding RMSD increment is hence 0.0089 Å. The close resemblance between the RMSD increments of the gold SL atoms (which are the origin of photoinduced motions) and the CH$_2$ units in the methylene lattice strongly supports a motion-based coupling for energy transport across the multiple components and high thermal conductivity along the chains.

Given that the Debye temperature is associated with the MSD (Eq. 2), we may consider an effective Debye temperature ($\Theta_{D,eff}$) of the methylene lattice along the out-of-plane direction approximated by

$$\frac{1}{\Theta_{D,eff}^2} = \left(\frac{\cos 30°}{\Theta_{D,\parallel}}\right)^2 + \left(\frac{\sin 30°}{\Theta_{D,\perp}}\right)^2,$$

where the nominal Debye temperatures associated with the intrachain ($\Theta_{D,\parallel}$) and interchain vdW directions ($\Theta_{D,\perp}$) are ~420 K and ~180 K for the MSDs of 0.017 Å$^2$ and 0.089 Å$^2$, respectively. Our calculated effective Debye temperature of ~290 K matches with what was suggested by Refs. (*26*, *52*) for ordered methylene chains. We further argue that the Debye temperature of 167 K considered in the low-temperature limit by Ref. (*54*) is not suitable for higher temperatures, where the corresponding RMSD at 290 K would be ~0.33 Å. This is much higher than a reasonable value considering the probed out-of-plane direction is still closer to the intrachain direction, which may have an RMSD of $\sqrt{0.017} = 0.13$ Å.

## 4. Thermal model for energy transport

We consider the time-dependent temperature increase of the gold SL $\Delta T_{Au}(t)$ based on the observed structural dynamics, with an instantaneous jump of 30 K followed by a 50-ps retention period and then an exponential decay with a nominal time constant of 460 ps and a long-time plateau of 2.5 K at 2.5 ns (Fig. S7A, red). The temperature offset is included to consider thermalization of the interfacial system on the long timescale. By treating the bonded sulfur atom as a boundary, the thermal coupling between the gold SL and the lowest methylene unit of the alkane chain is described by

$$K \frac{\partial \Delta T_{SAM}(z,t)}{\partial z}\bigg|_{z=0} = -\sigma[\Delta T_{Au}(t) - \Delta T_{SAM}(z,t)],$$

where $\sigma$ is the thermal boundary conductance, $K$ is the effective thermal conductivity along the surface normal direction, and $T_{SAM}(z,t)$ is the effective temperature increase of the $j$-th methylene unit from the gold surface with $z = (j-1) \cdot 0.11$ nm. The following one-dimensional thermal diffusion equation is used for the finite-difference numerical computation with the Crank–Nicolson algorithm,

$$\frac{\partial^2 \Delta T_{SAM}(z,t)}{\partial z^2} - \frac{c}{K}\frac{\partial \Delta T_{SAM}(z,t)}{\partial t} = 0$$

where $c = 2.96$ J cm$^{-3}$ K$^{-1}$ is the volumetric heat capacity of the methylene lattice (*29*). The initial and boundary conditions are

$$\Delta T_{SAM}(z,0) = 0, \quad \text{and} \quad \frac{\partial \Delta T_{SAM}(z,t)}{\partial z}\bigg|_{z=n\cdot 0.127\text{nm}} = 0,$$

the latter of which indicates no energy flow across the SAM–vacuum interface. Thus, a weighted average is obtained by

$$\Delta \bar{T}_{SAM}(t) = \Sigma_m P_m \cdot \Delta T_{SAM}(z_m, t).$$

## 5. Negligible impact of transient electric field effects on UED dynamics

Recent reflection UED studies have unequivocally shown the negligible impact of photoinduced transient electric fields (TEFs) on the observed structural dynamics near sample surfaces at low-to-moderate excitation fluences (*23, 35, 38, 39, 49, 50, 54, 55*). In many cases, the TEFs were in fact not noticeable. Even when TEFs were observed in a study of metal surfaces, their effects on the structural dynamics were shown to be negligible (*23*). A similar conclusion can be reached based on the comparison of the intensity changes of the shadow-edge region and the SAM diffraction spots shown in Fig. S9. Further definite evidence is the observation of an ultrafast gold SL dynamics acquired at a higher takeoff angle compared to those of the slower HG and SAM dynamics recorded at low takeoff angles. The expected outcomes from the TEFs, if real, would be a fastest response for the HG diffraction streaks as they are very close to the shadow edge, a slower development of changes for the SAM Bragg spots, and a much slowed and reduced influence for the gold SL diffractions farthest from the shadow edge. Thus, we further confirm the confidence of using reflection UED to study structural dynamics at low-to-moderate levels of photoexcitation.

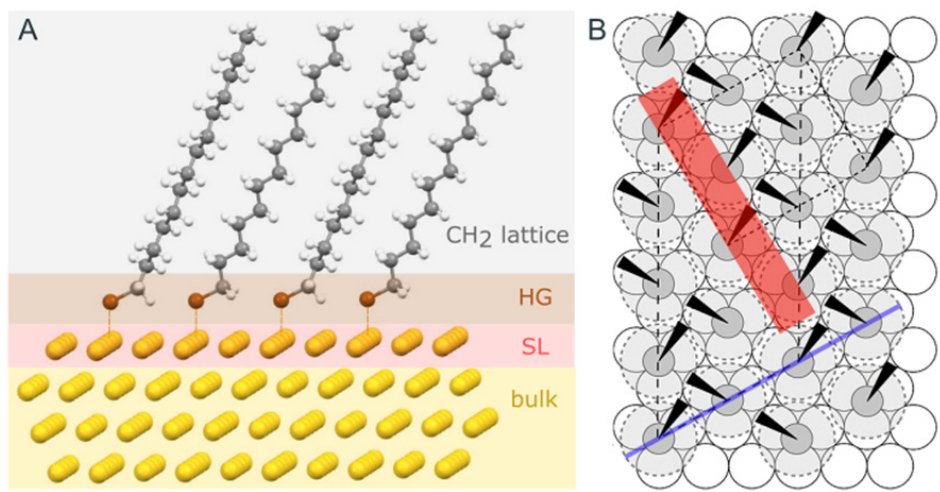

**Fig. S1. Schematic of the Au(111)–SAM system.** (A) Another view of a monolayer of *n*-alkanethiolates with 14-carbon chains chemically bonded to Au(111) (carbon in gray, hydrogen in white, sulfur in brown, and gold in yellow). The chains exhibit two distinct twist angles that alternate in the monolayer (*16*). (B) Top view of the ordered SAM formed on Au(111), adapted from Ref. (*16*). Sulfur atoms (dark gray) occupy three-fold hollow sites of the gold lattice (white circles), showing a predominantly ($\sqrt{3}\times\sqrt{3}$)R30° adlayer. Light gray dashed circles mark the approximate surface footprint of each alkane chain, and dark wedges show the orientation of the chain's CCC plane projected onto the surface. The red band and the blue line indicate the horizontal cuts as shown in Fig. 1A and fig. S1A, respectively.

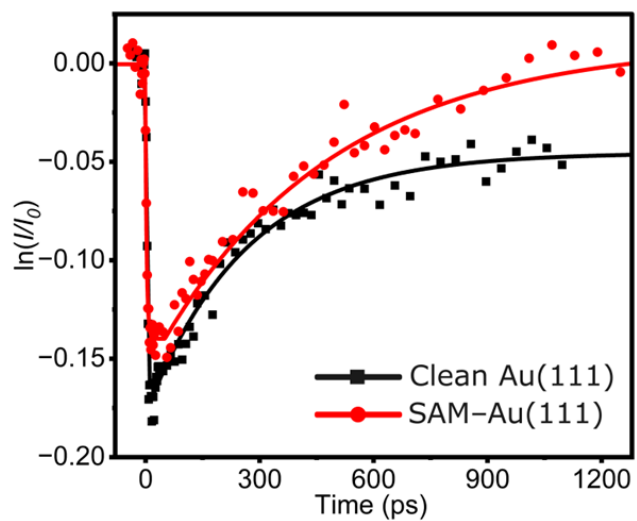

**Fig. S2. Time-resolved diffraction intensity changes of clean (black) and SAM-bonded (red) Au(111) SLs**, which reach the maximum reductions of ~17% and ~14%, respectively, at a fluence of 4.4 mJ/cm$^2$.

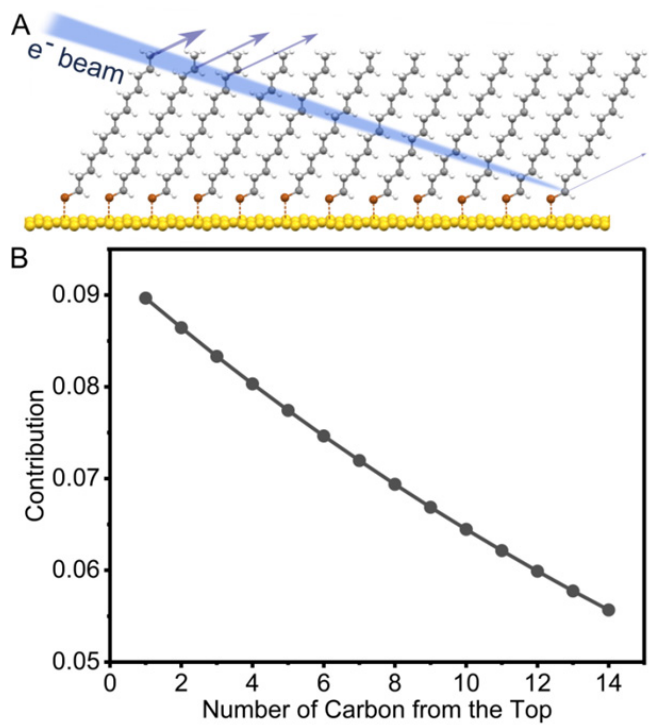

**Fig. S3. Depth-dependent scattering cross section of the SAM film.** (A) Schematic of reduced scattered electron contributions (indicated by thinner arrows) from lower $CH_2$ units along the methylene chain of a SAM. (B) Weighted contributions to the diffraction signals from the carbon atoms numbered along the chain from the top. The example given here is for the 1-tetradecanethiol SAM.

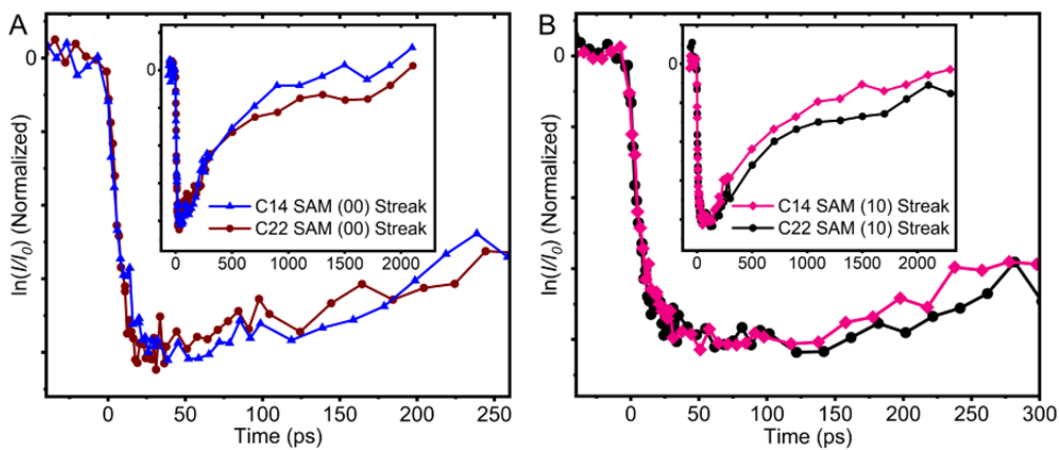

**Fig. S4. Comparison of the time-dependent intensity changes** of (A) the center streak and (B) the side streak from SAMs with two select chain lengths. The insets show the longer timescale. The comparable rise times of ~13 ps and the similar retention times observed are consistent with the physical picture that the HGs are bonded with the gold SL, which act as the energy source, and that the coupling of atomic motions and energy transfer is essentially insensitive to the alkane chain length if the overlayer thickness is sufficient.

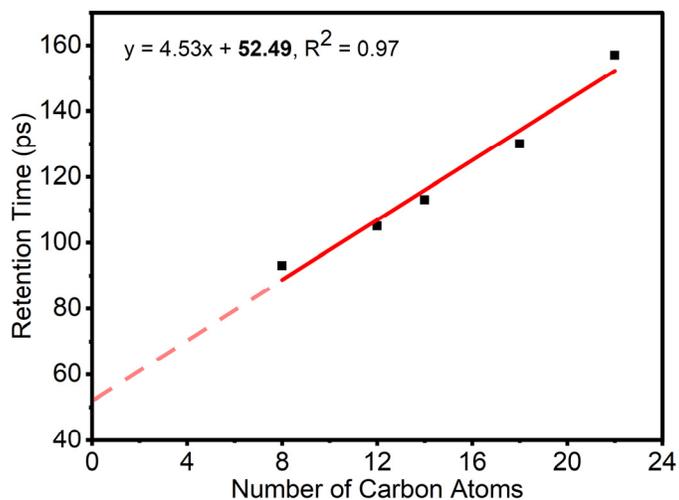

**Fig. S5. Length-dependent retention time for the methylene lattice dynamics.** The extrapolation of a linear fit (red line) of the observed retention times for different chain lengths yields an intercept of ~52 ps at zero length, which matches well with the retention period observed for the gold SL (Fig. 2, A and B, red).

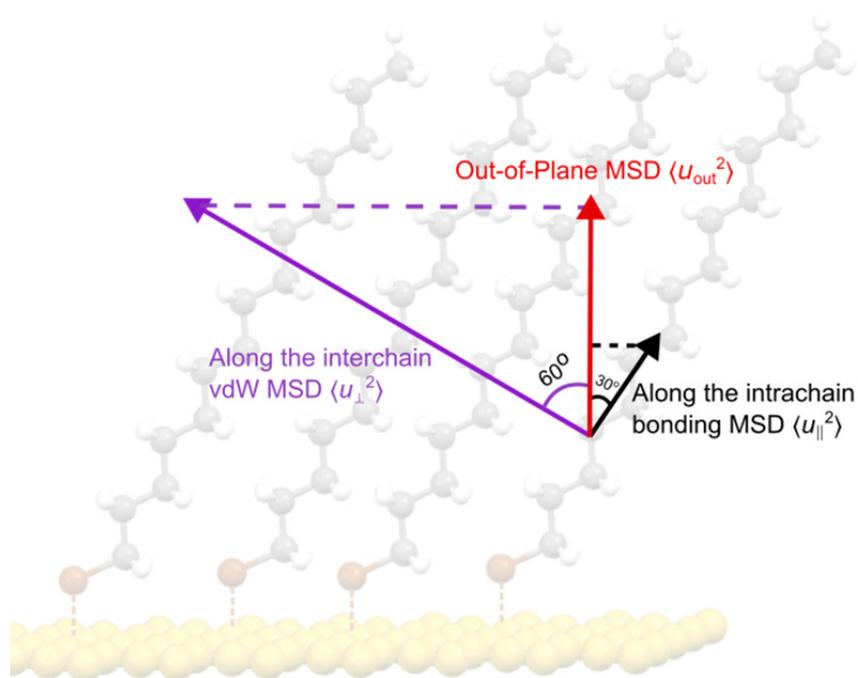

**Fig. S6. Schematic for the geometry considered to calculate the MSD and effective Debye temperature of the methylene lattice along the probing direction**, which is normal to the surface (red arrow) and ~30° away from the all-trans intrachain direction (black arrow). The purple arrow perpendicular to the bonded chains indicates the interchain vdW direction. The vectors are drawn proportionally to reflect the magnitudes of the MSDs along the specified directions.

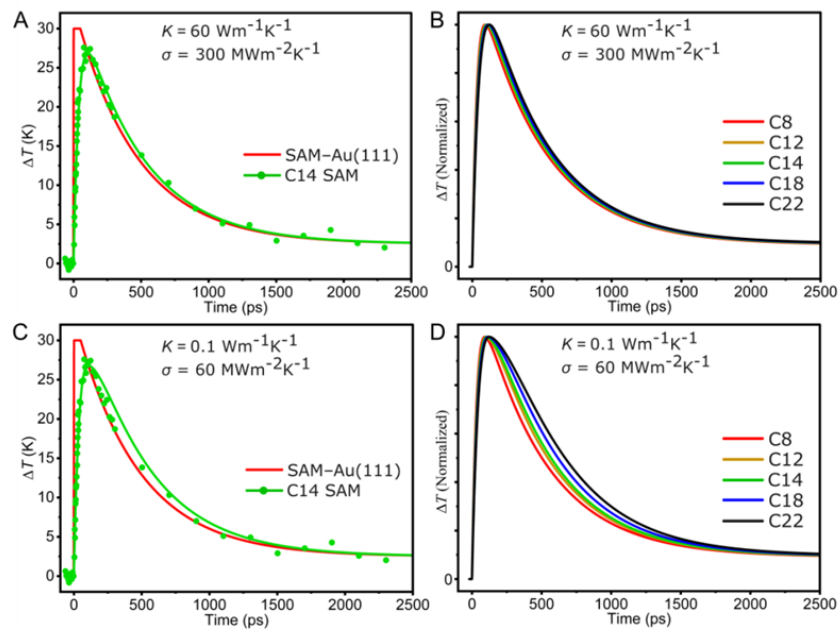

**Fig. S7. Time-dependent dynamics of the methylene lattice assessed by the thermal model for energy relaxation.** (A and C) Comparison of the experimentally-deduced temperatures of a C14 SAM (green dots) and the theoretical time-dependent curve obtained with (A) $\sigma$ = 300 MW m$^{-2}$ K$^{-1}$ and $K$ = 60 W m$^{-1}$ K$^{-1}$ and (C) $\sigma$ = 60 MW m$^{-2}$ K$^{-1}$ and $K$ = 0.1 W m$^{-1}$ K$^{-1}$. The red curve shows the experimentally-determined temperature profile used for the gold SL. (B and D) Theoretical curves of temperature evolution with increasing chain length obtained with $K$ and $\sigma$ values of the panel A and C respectively.

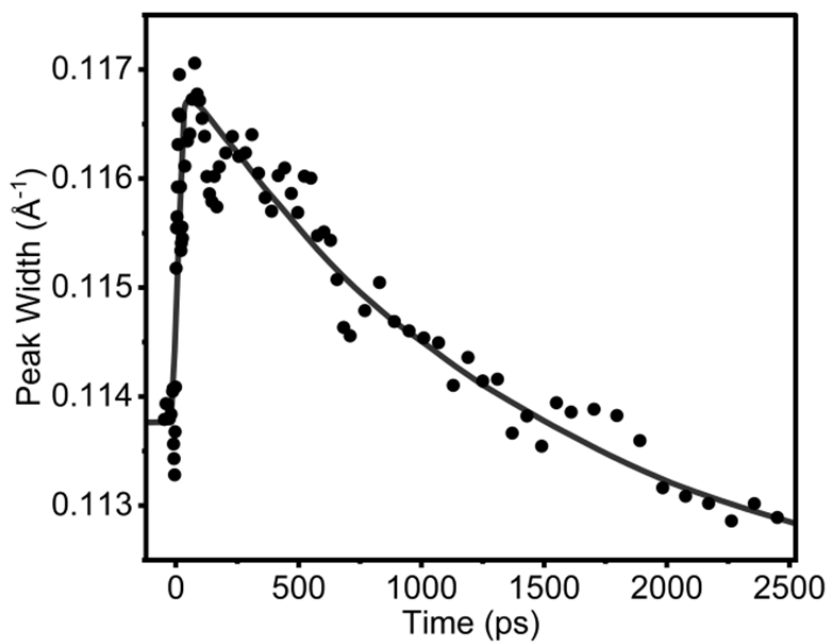

**Fig. S8. Time-dependent evolution of the horizontal width of the gold SL diffraction.** The data points are shown with 5-point moving average. The solid curve is a guide to the eye.

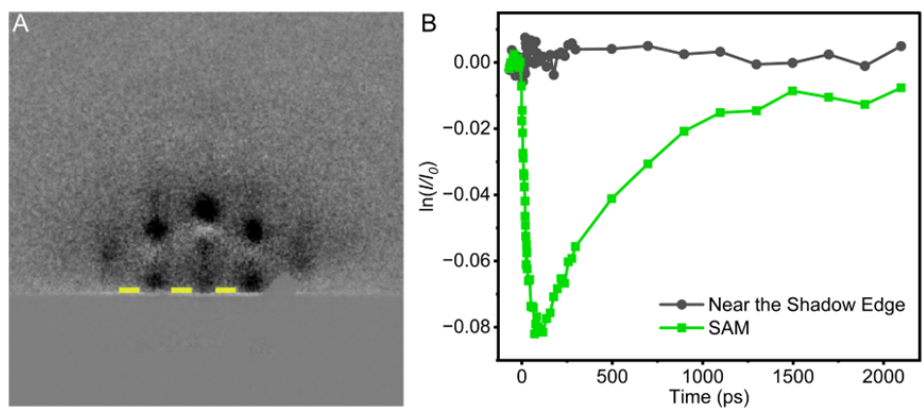

**Fig. S9. Comparison of diffraction changes caused by transient electric fields vs. by structural dynamics.** (A) Diffraction difference image recorded at 50 ps, with boosted contrast to show the small changes near the shadow edge. The yellow rectangles between diffraction features are used to track the intensity change. (B) Time-dependent intensity changes of the SAM dynamics and the shadow-edge region.